\documentclass[10pt,conference]{IEEEtran}

\IEEEoverridecommandlockouts

\IEEEaftertitletext{\vspace{-0\baselineskip}}
\usepackage{cite}
\usepackage{amsmath,amssymb,amsfonts}
\usepackage{algorithmic}
\usepackage{graphicx}
\usepackage{textcomp}
\usepackage{xcolor}
\usepackage{graphicx}  
\usepackage{booktabs} 
\usepackage{adjustbox}  
\usepackage{algorithm}
\usepackage{algorithmic}
\usepackage{tikz}
\usepackage{pgfplots}
\pgfplotsset{compat=1.18}

\newcommand{\black}[1]{{\textcolor[rgb]{0, 0, 0}{#1}}}

\usetikzlibrary{arrows.meta,positioning}

\begin{document}
\title{{Semantic-Aided Iterative Decoding for Uplink Non-Orthogonal Transmission}}
\author{
Wenjing Wei, Chentao Yue, Branka Vucetic, Yonghui Li\\
School of Electrical and Computer Engineering, The University of Sydney, Australia\\
E-mail: \{wenjing.wei, chentao.yue, branka.vucetic, yonghui.li\}@sydney.edu.au

\thanks{The work of Chentao Yue was supported by ARC DECRA under Grant DE250101332. Code available: https://github.com/Wenjing79/NOMA-LLM.}
}

\maketitle
\begin{abstract}
This paper proposes semantic-aided iterative decoding
(Sem-IR) for uplink non-orthogonal transmission of a shared
natural-language source. $K$ users each hold one segment of a
common sentence and superimpose low-density parity-check (LDPC) coded transmissions over an
additive white Gaussian noise (AWGN) channel. At the base station,
an iterative elementary signal estimator (ESE) and $K$ parallel
LDPC decoders progressively cancel inter-user interference.
\black{As high-power users pass both parity and language-plausibility
checks earlier, their decoded bytes form a reliable linguistic
prefix for the remaining users;} a fine-tuned ByT5
byte-level language model exploits this prefix to predict byte
posteriors for the unconverged user. The byte posteriors are
marginalized to bit-level log-likelihood ratios and convex-combined with the LDPC posteriors inside the iterative loop. The resulting feedback closes the
loop between the language model and the physical-layer iteration.
Simulations show that Sem-IR outperforms orthogonal time-division
access (TDMA) and the same NOMA receiver without semantic feedback
in block error rate (BLER), {
yielding an order-of-magnitude reduction over NOMA at $8$\,dB.}

\end{abstract}

\vspace{-0.5em}
\section{Introduction}
\vspace{-0.3em}
Semantic communication has emerged as a new paradigm that
prioritizes meaningful information delivery over strict bit-level
accuracy~\cite{10965715,10818533}. {This perspective aligns with
6G visions, where communication supports task execution, control,
and inference under stringent bandwidth, latency, and reliability
constraints~\cite{10054381}, motivating receivers that exploit
prior knowledge, context, and task goals to preserve meaning
under reduced overhead.}

{Many semantic communication schemes adopt deep
joint source--channel coding (JSCC), in which neural
encoders/decoders directly map sources to channel
symbols~\cite{10824931,10770606}. Recently, a progressive JSCC framework improves image super-resolution in noisy channels, and D$^2$-JSCC introduces adaptive density modeling and joint source--channel rate control to improve semantic reconstruction \cite{10973232,10845799}. While JSCC offers graceful
degradation under varying SNR, its neural transceivers are
typically trained for specific modalities, tasks, and channel
models, and replace standardized physical-layer modules entirely,
limiting generalization and deployment.}

{An alternative line of work preserves the standardized
channel code and uses a large language model (LLM) at the
receiver as a context-aware error
corrector~\cite{10822884,11432683,Wang2026CLSEC}.
Representative schemes encode short text segments with classical
codes and apply a generative model after channel decoding to
repair residual errors~\cite{11432683}. While such methods improve
semantic-level fidelity over physical-layer decoding alone, the LLM output reshapes
the final hard decision but is never fed back to refine the
physical-layer estimation.}

A complementary opportunity arises in multi-user uplink scenarios
typified by massive machine-type communications (mMTC) and
distributed sensing, where many devices share spectrum to upload
fragmentary observations to a common
aggregator~\cite{ShahabGFNOMA2020,Jabbarvaziri2021HARQ}.
{Power-domain non-orthogonal multiple access (NOMA) is widely
adopted in such systems for its massive-connectivity gain.}
When the uploaded payloads are textual; for instance,
distributed nodes reporting fragments of a common event
description, or collaborative agents transmitting a structured
report~\cite{Peng2025LLMDiSAC}, streams from different users are
linguistically correlated through the shared context. This
correlation, latent in the source rather than in the channel,
remains untapped by conventional NOMA receivers. The
interleave-division multiple access (IDMA)
framework~\cite{1452830,6092790} {provides a practical iterative receiver for power-domain
NOMA, but likewise treats user payloads as independent
random bit streams.} Closing this gap requires a receiver that
exploits cross-user source structure
for multi-user detection.

{
This paper proposes a semantic-aided iterative receiver (Sem-IR) for
uplink non-orthogonal transmission of a shared natural-language
source. We consider $K$
users (e.g., distributed sensing nodes or collaborative
agents) each holding one contiguous byte segment of a common
sentence. Each user independently encodes its segment with a
low-rate quasi-cyclic LDPC (QC-LDPC) code \cite{Fossorier2004QCLDPC}, and transmits over a shared additive white Gaussian
noise (AWGN) channel under a geometric power profile. At the base
station, an iterative elementary signal estimator (ESE) and $K$
parallel LDPC decoders progressively cancel inter-user
interference.
\black{The higher-power users pass the parity and language-plausibility
checks earlier, so their accepted bytes form a reliable linguistic
prefix for the remaining users.}
Our proposed receiver employs a fine-tuned byte-level language model to use this prefix to
predict byte posteriors for \black{unaccepted} users. The
posteriors are marginalized to bit-level log-likelihood ratios
(LLRs), gated by a confidence threshold, and convex-combined with
the LDPC a-posteriori LLRs before re-entering the iterative receiver loop. Once a user is recovered with the help of the semantic prior, the corresponding LLR injection is retained until decoding terminates.
Simulations show that the proposed scheme outperforms orthogonal serial transmission (TDMA) and the same
NOMA receiver without semantic feedback in terms of block error
rate, with the
gap to NOMA reaching one order of magnitude at $8$\,dB.}

The rest of the paper is organized as follows:
Sections~\ref{sec:system_model}--\ref{sec:conclusion} present the
system model, Sem-IR, simulation results, and conclusion, respectively.

\vspace{-0.8em}
\section{System Model} 
\vspace{-0.3em}
\label{sec:system_model}
 \begin{figure*}[!t]  
  \centering
  \includegraphics[width=0.9\textwidth]{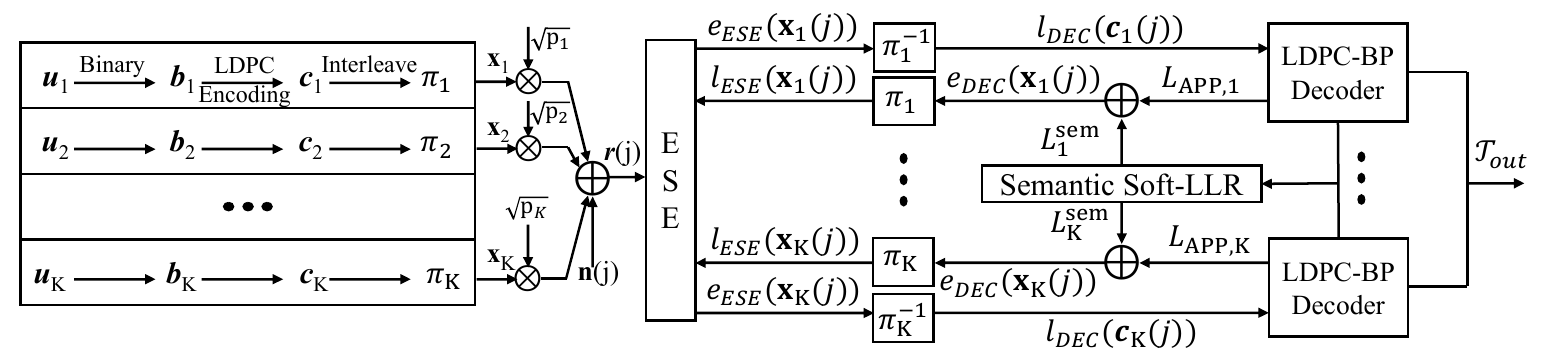}  
  \vspace{-0.5em}
  \caption{Overall Architecture of {the proposed semantic-aided
iterative receiver for LDPC-encoded uplink non-orthogonal
transmission}}
  \label{fig:idma_bart_system}
  \vspace{-1.5em}
\end{figure*}

We consider a system that conveys a single coherent natural-language sentence
over an AWGN channel through superimposed
LDPC-coded layers {transmitted by
$K$ uplink users}.
{Each user holds one of $K$ contiguous byte segments of the
sentence and encodes it as one layer of the superposition.}
At the receiver, the $K$ layers are jointly detected and decoded by
an iterative ESE and $K$ parallel BP
decoders, augmented by a byte-level semantic prior when
conventional BP decoding fails.  {The overall architecture is illustrated in Fig.\ 1.} This section formalises the
transmission structure and the conventional iterative ESE--DEC receiver.

\vspace{-0.3em}
\subsection{Transmission Structure}
\vspace{-0.3em}
\black{
Let $\mathcal{T}$ denote the source sentence of $N$~UTF-8 bytes.
We pad $\mathcal{T}$ to a multiple of $K$ when necessary and partition the
result into $K$ equal-length byte segments $\{\boldsymbol{\mu}_k\}_{k=1}^{K}$
with $\boldsymbol{\mu}_k \in \mathbb{F}_{2^8}^{L}$ and $L = \lceil N/K \rceil$.
Each segment is unpacked into its binary representation
$\boldsymbol{b}_k \in \mathbb{F}_2^{M}$, $M = 8L$,
and encoded by a rate-$R$ QC-LDPC code into a codeword
$\boldsymbol{c}_k \in \mathbb{F}_2^{n}$, $n = M / R$.}

\black{
A segment-specific random interleaver $\pi_k$ is applied to $\boldsymbol{c}_k$,
and the result is BPSK-modulated to
$\mathbf{x}_k = (x_k(1), \ldots, x_k(n)) \in \{\pm 1\}^{n}$,
$x_k(j) = 1 - 2c_k\bigl(\pi_k(j)\bigr)$.
The interleavers $\{\pi_k\}_{k=1}^{K}$ are drawn independently and uniformly
at random from the symmetric group on $\{1, \ldots, n\}$, and they serve as
layer signatures that distinguish the $K$ layers at the receiver.
}

The $K$ modulated layers are weighted and superimposed.
{Following the geometric
power allocation typical of power-domain NOMA}, layer $k$ carries transmit power $p_k$
allocated according to the geometric profile
\begin{equation}
  p_k \;=\; \frac{\rho^{\,K-k}}{\sum_{u=1}^{K} \rho^{\,K-u}},
  \qquad k = 1, \ldots, K,
  \label{eq:power}
\end{equation}
where $\rho \ge 1$ controls the power gap between adjacent layers and
$\sum_{k=1}^{K} p_k = 1$.

{
The received signal at position $j \in \{1, \ldots, n\}$ is
\begin{equation}
  r(j) \;=\; \sum_{k=1}^{K} \sqrt{p_k}\, x_k(j) \;+\; w(j),
  \label{eq:rx}
\end{equation}
where $w(j) \sim \mathcal{N}(0, \sigma^{2})$ is i.i.d.\ AWGN with
$\sigma^{2} = N_0 / 2$. We define $\mathrm{SNR} = 1 / \sigma^{2}$.
}

\vspace{-0.3em}
\subsection{Receiver Structure}\label{sec:idma_skeleton}
\vspace{-0.3em}
\black{
The receiver runs at most $T_{\max}$ outer iterations, each consisting of
one ESE pass and $K$ parallel BP decodes.
We index outer iterations by $t = 1, 2, \ldots$ and use superscript
$(t)$ for any quantity at iteration~$t$.
}

\black{\subsubsection{Elementary Signal Estimator (ESE)}
At iteration~$t$, the ESE treats interference from the other $K-1$ layers
as Gaussian and produces, for position $j$ and layer $k$, the extrinsic
LLR
\begin{equation}
  e_{\mathrm{ESE}}^{(t)}(x_k(j))
  \;=\; \frac{2 \sqrt{p_k}\bigl(r(j) - E[\xi_k(j)]\bigr)}
             {\mathrm{Var}(\xi_k(j))},
  \label{eq:ese}
\end{equation}
where $\xi_k(j) = r(j) - \sqrt{p_k}\, x_k(j)$ has moments
\begin{align}
  E[\xi_k(j)]
    &= \sum_{k' \neq k} \sqrt{p_{k'}}\, E\!\left[x_{k'}(j)\right],
  \label{eq:xi_mean}\\
  \mathrm{Var}(\xi_k(j))
    &= \sum_{k' \neq k} p_{k'}\, \mathrm{Var}\!\left(x_{k'}(j)\right)
       \;+\; \sigma^{2}.
  \label{eq:xi_var}
\end{align}
The soft moments $E[x_{k'}(j)]$ and $\mathrm{Var}(x_{k'}(j))$ are
fed back from the DEC at iteration $t-1$ via~\eqref{eq:mean} and \eqref{eq:var}.
}

\subsubsection{Decoder (DEC)}
The ESE output for layer $k$ is de-interleaved by $\pi_k^{-1}$ to form the
channel LLR
$\boldsymbol{\ell}_{\mathrm{DEC},k}^{(t)} \in \mathbb{R}^{n}$
in the codeword domain.
The LDPC decoder runs up to $T_{\mathrm{bp}}$ rounds of BP
on $\boldsymbol{\ell}_{\mathrm{DEC},k}^{(t)}$ and produces a hard decision
$\hat{\boldsymbol{c}}_k^{(t)}$ together with the a posteriori LLR
$\boldsymbol{L}_{\mathrm{APP},k}^{(t)}$.
The decoder extrinsic returned to the ESE is
\begin{equation}
  e_{\mathrm{DEC},k}^{(t)}(j)
  \;=\; L_{\mathrm{APP},k}^{(t)}(j)
        \;-\; \ell_{\mathrm{DEC},k}^{(t)}(j),
  \quad j = 1, \ldots, n.
  \label{eq:dec_ext}
\end{equation}

After re-interleaving by $\pi_k$, $e_{\mathrm{DEC},k}^{(t)}$ becomes the
a priori LLR $\ell_{\mathrm{ESE},k}^{(t+1)}$ used by the next ESE pass.
\black{Layer $k$ is declared parity-valid when its hard decision satisfies
all parity checks,
$\mathbf{H}\hat{\boldsymbol{c}}_k^{(t)\mathsf{T}}
\equiv \boldsymbol{0}$, where $\mathbf{H}$ is the LDPC parity-check
matrix.}
\black{A zero syndrome establishes codeword validity but does not exclude
an incorrect valid codeword; final runtime acceptance therefore also
requires the language-plausibility test in Section~\ref{sec:byt5}.}

\subsubsection{Soft Symbol Update}
Bit-level LLRs are converted into BPSK soft moments via
\begin{align}
  E\!\left[x_k(j)\right]
    &= \tanh\!\Bigl(\tfrac{1}{2}\, \ell_{\mathrm{ESE},k}^{(t+1)}(j)\Bigr),
  \label{eq:mean}\\
  \mathrm{Var}\!\left(x_k(j)\right)
    &= 1 - \bigl(E[x_k(j)]\bigr)^{2}.
  \label{eq:var}
\end{align}

To suppress oscillation of the soft moments under strong
inter-layer interference, the a priori LLR
$\ell_{\mathrm{ESE},k}^{(t+1)}(j)$ in \eqref{eq:mean} is usually damped
across iterations as
\begin{equation}
   \!\tanh^{\!-1}\!\!\left(\!
      \beta \tanh\!\!\Big(\tfrac{\ell_{\mathrm{ESE},k}^{(t+1)}(j)}{2}
        \Big)
     \! + \!(1\!-\!\!\beta)\tanh\!\!\!\!\ \Big(\tfrac{\tilde{\ell}_{\mathrm{ESE},k}^{(t)}(j)}{2}
        \Big)\!\!
    \right),
  \label{eq:damping}
\end{equation}
with damping factor $\beta\in(0,1]$.

{
At low to moderate $\mathrm{SNR}$, however, the ESE--DEC loop alone
frequently stalls, leaving one or more
layers unconverged.
Section~\ref{sec:receiver} introduces a byte-level semantic prior that is
injected into $\boldsymbol{L}_{\mathrm{APP},k}^{(t)}$ for
 \black{unaccepted} layers, breaking the loop out of such local optima.
}

\vspace{-0.5em}
\section{Semantic-Assisted Iterative Receiver} \label{sec:receiver}
\vspace{-0.3em}

\begin{figure}
  \centering
  \includegraphics[width=0.4\textwidth]{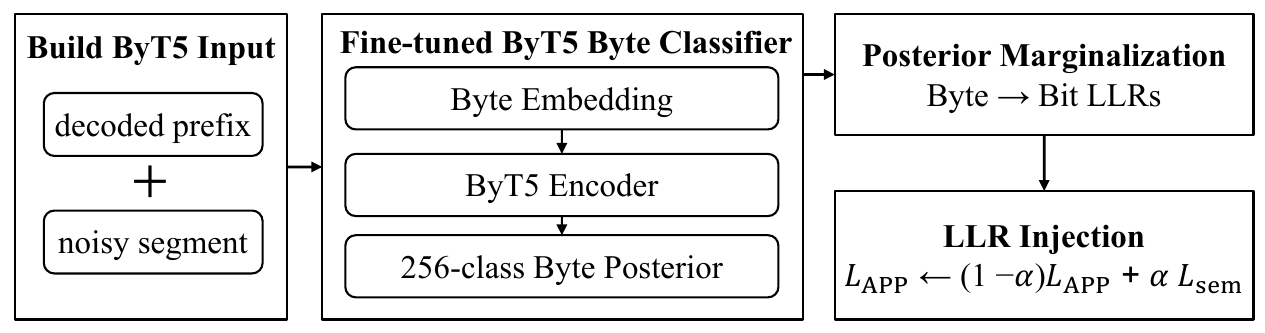}
  \vspace{-0.5em}
  \caption{ Semantic Prior Module.}
  \label{fig:idma_byt5_system}
  \vspace{-1.25em}
\end{figure}

{
When the ESE--DEC receiver of Section~\ref{sec:idma_skeleton} stalls, we
invoke a byte-level semantic prior on the lowest-index \black{unaccepted} layer. The structure of semantic prior module is showed in Fig. \ref{fig:idma_byt5_system}.
}

\vspace{-0.3em}
\subsection{Byte-Level Semantic Prior}
\label{sec:byt5}
\vspace{-0.3em}

\subsubsection{Model Forward Pass}

{
At the end of a specific outer iteration $t$, we identify the lowest-index
\black{segment $k\notin\Omega_{\mathrm{acc}}^{(t)}$, where
$\Omega_{\mathrm{acc}}^{(t)}$ denotes the segments passing both the
parity and language-plausibility checks, as formalized in
\eqref{eq:semantic_acceptance}. The $k-1$ preceding segments belong to
$\Omega_{\mathrm{acc}}^{(t)}$ and form the accepted decoded prefix. A
parity-valid segment that fails the language-plausibility test remains
unaccepted and can therefore be selected for semantic repair.}
Segments with index $>k$ are excluded even when \black{accepted}, so that the
prefix supplied to the language model remains contiguous.
The noisy byte estimate $\hat{\boldsymbol{\mu}}_k^{(t)}\in
\mathbb{F}_{2^8}^{L}$ is repacked from the BP information-bit hard
decisions of segment $k$ at iteration $t$.
As $t$ grows, $k$ advances and the prefix expands until
the full sentence is recovered.
}

{
The byte-level semantic prior is realized by the pre-trained ByT5 encoder~\cite{xue-etal-2022-byt5} followed by a linear classification head over $\mathbb{F}_{2^8}$.
The encoder input concatenates the prefix and the
noisy segment estimate,
\begin{equation}
  \boldsymbol{u}_k^{(t)} = [
    \black{\boldsymbol{\mu}_{1:k-1}}\ \ \ \ 
    \hat{\boldsymbol{\mu}}_k^{(t)}].
  \label{eq:byt5_input}
\end{equation}
Let $\{q_{k,1},\ldots,q_{k,L}\}$ denote the positions of
$\hat{\boldsymbol{\mu}}_k^{(t)}$ within $\boldsymbol{u}_k^{(t)}$.
The encoder, parameterized by $\theta$, produces hidden states
$\boldsymbol{H}_k^{(t)}= f_{\mathrm{enc}}(\boldsymbol{u}_k^{(t)};\theta)\in{\mathbb{R}^{|\boldsymbol{u}_k^{(t)}|\times d}}$
of hidden dimension $d$, whose $q$-th row $\boldsymbol{h}_{k,q}^{(t)}$}

The head produces logits
$\boldsymbol{z}_{k,i}^{(t)}=
\mathbf{W}\,
{\boldsymbol{h}_{k,q_{k,i}}^{(t)}}
+\boldsymbol{a}\in\mathbb{R}^{256}$ at each noisy byte
position $i\in\{1,\ldots,L\}$, with weight matrix
$\mathbf{W}\in\mathbb{R}^{256\times d}$ and bias
$\boldsymbol{a}\in\mathbb{R}^{256}$.
The byte posterior over $\mathbb{F}_{2^8}$ is
\begin{equation}
  p_{k,i}(b\mid
   \black{\boldsymbol{\mu}_{1:k-1}},
    \hat{\boldsymbol{\mu}}_k^{(t)})
  = \frac{\exp\bigl(z_{k,i}^{(t)}(b)\bigr)}
         {\sum_{b'\in\mathbb{F}_{2^8}}
           \exp\bigl(z_{k,i}^{(t)}(b')\bigr)}.
  \label{eq:byte_post}
\end{equation}
where $z_{k,i}^{(t)}(b)$ denotes the $b$-th entry of
$\boldsymbol{z}_{k,i}^{(t)}$ .

\black{For a parity-valid segment, the same posterior also assesses the
plausibility of its decoded bytes. Define the average and worst-byte
log-probability scores and the accepted set as
\begin{small}
\begin{equation}
\begin{aligned}
&s_{k,\mathrm{avg}}^{(t)}
  = \frac{1}{L}\sum_{i=1}^{L}
     \log p_{k,i}\!\left(\hat{\mu}_{k,i}^{(t)}\mid
    \boldsymbol{\mu}_{1:k-1}^{(t)},
     \hat{\boldsymbol{\mu}}_k^{(t)}\right),\\
&s_{k,\min}^{(t)}
  = \min_{1\le i\le L}
     \log p_{k,i}\!\left(\hat{\mu}_{k,i}^{(t)}\mid
    \boldsymbol{\mu}_{1:k-1}^{(t)},
     \hat{\boldsymbol{\mu}}_k^{(t)}\right),\\
&\Omega_{\mathrm{acc}}^{(t)}
  \!=\! \left\{k:\,
    \!\! \mathbf H\hat{\boldsymbol c}_k^{(t)\mathsf T}\!=\!\boldsymbol 0,
     s_{k,\mathrm{avg}}^{(t)}\ge\tau_{\mathrm{avg}},
     s_{k,\min}^{(t)}\ge\tau_{\min}\right\}.
\end{aligned}
\label{eq:semantic_acceptance}
\end{equation}
\end{small}
Only segments in $\Omega_{\mathrm{acc}}^{(t)}$ are used as linguistic
context or accepted for receiver termination.}

{
\subsubsection{Training}
The model is fine-tuned by minimizing the per-position cross-entropy
\begin{equation}
  \mathcal{L}_{\mathrm{CE}}
  = -\frac{1}{L}\sum_{i=1}^{L}
    \log p_{k,i}(\mu_{k,i}\mid
      \boldsymbol{\mu}_{1:k-1},\hat{\boldsymbol{\mu}}_k),
  \label{eq:ce_loss}
\end{equation}
where $\mu_{k,i}\in\mathbb{F}_{2^8}$ denotes the $i$-th byte of
$\boldsymbol{\mu}_k$.
Training samples follow the deployment protocol:
$\boldsymbol{\mu}_{1:k-1}$ is taken from the ground-truth sentence and
$\hat{\boldsymbol{\mu}}_k$ from the BP hard decisions of the
lowest-index erroneous segment in a simulated NOMA pass, so the training
input distribution matches~\eqref{eq:byt5_input}.
}
\vspace{-0.3em}
\subsection{Bit-Level Semantic LLR and Confidence-Gated Injection}\label{sec:soft_llr}
\vspace{-0.3em}

\subsubsection{Byte-to-Bit Marginalization}
The LDPC decoder operates on bit-level LLRs, whereas
\eqref{eq:byte_post} is byte-valued. For each information-bit
position $\ell\in\{1,\ldots,M\}$ of segment $k$, write
$\ell=8(i-1)+j+1$ with byte index $i\in\{1,\ldots,L\}$ and
intra-byte index $j\in\{0,\ldots,7\}$. Marginalizing
\eqref{eq:byte_post} over all byte values whose $j$-th bit equals
$\beta\in\{0,1\}$ yields the bit-level semantic log-posterior
\begin{equation}
  \lambda^{\mathrm{sem}}_{k,\ell}(\beta)
  = \log\!\!\sum_{\substack{b\in\mathbb{F}_{2^8}\\ \mathrm{bit}_j(b)=\beta}}
        p_{k,i}\bigl(b\mid
                           \boldsymbol{\mu}_{1:k-1},
                          \hat{\boldsymbol{\mu}}_k^{(t)}\bigr),
  \label{eq:bit_sem_logpost}
\end{equation}
where $\mathrm{bit}_j(b)$ denotes the $j$-th bit of $b$ in big-endian
order. The corresponding bit-level semantic LLR is
\begin{equation}
  L^{\mathrm{sem}}_{k,\ell}
  = \lambda^{\mathrm{sem}}_{k,\ell}(0)-\lambda^{\mathrm{sem}}_{k,\ell}(1).
  \label{eq:bit_sem_llr}
\end{equation}
\vspace{-1em}
\black{\subsubsection{Confidence-Gated Fusion}
At each ByT5 invocation on segment $k$, we form the activation set
\begin{equation}
  \mathcal{A}_k
  = \bigl\{\,\ell\in\{1,\ldots,M\}:
    \bigl|L^{\mathrm{sem}}_{k,\ell}\bigr|\ge\gamma\bigr\},
  \label{eq:active_set}
\end{equation}
where $\gamma>0$ is a confidence threshold. For $\ell\in\mathcal{A}_k$
we freeze the semantic LLR computed at this invocation,
$\widetilde{L}^{\mathrm{sem}}_{k,\ell}\triangleq L^{\mathrm{sem}}_{k,\ell}$,
and inject it into the LDPC a posteriori LLR by convex fusion
\begin{equation}
  \boldsymbol{L}_{\mathrm{APP},k}^{(t)}(\ell)
  \;\leftarrow\;
  (1-\alpha)\boldsymbol{L}_{\mathrm{APP},k}^{(t)}(\ell)
  +\alpha\,\widetilde{L}^{\mathrm{sem}}_{k,\ell},
  \quad \ell\in\mathcal{A}_k,
  \label{eq:fusion}
\end{equation}
with mixing weight $\alpha\in(0,1]$; $\alpha=1$ corresponds to fully
replacing $\boldsymbol{L}_{\mathrm{APP},k}^{(t)}(\ell)$ by $\widetilde{L}^{\mathrm{sem}}_{k,\ell}$, while $\alpha<1$ retains a fraction of the channel
evidence. Bits with $|L^{\mathrm{sem}}_{k,\ell}|<\gamma$ are
discarded and the corresponding APP-LLRs are left unchanged. Note
that the activation set $\mathcal{A}_k$ and the frozen values
$\{\widetilde{L}^{\mathrm{sem}}_{k,\ell}\}_{\ell\in\mathcal{A}_k}$
are determined once per ByT5 invocation.
}

{In practice, $\alpha$ is set close to one. The reason is that when segment $k$ has not converged
after the iterative ESE--DEC loop, the residual inter-layer
interference on its bits has not been cancelled. The
resulting $\boldsymbol{L}_{\mathrm{APP},k}^{(t)}$ is therefore unreliable in both magnitude and sign. a large $\alpha$ lets the semantic prior
dominates, while the gating
\eqref{eq:active_set} ensures that only bits with reliable semantic evidence are overwritten.}

\vspace{-0.3em}
\subsection{Iterative Schedule}

\vspace{-0.3em}

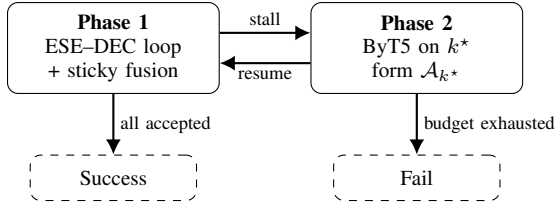
\begin{figure}[t]
\centering
\scriptsize
\setlength{\baselineskip}{0.92\baselineskip}
\begin{tikzpicture}[
  node distance=8mm and 12mm,
  font=\footnotesize,
  >={Latex[length=2mm,width=2mm]},
  block/.style={rectangle, draw, rounded corners,
                minimum height=12mm, minimum width=28mm,
                align=center, inner sep=3pt},
  term/.style={rectangle, draw, dashed, rounded corners,
               minimum height=6mm, minimum width=22mm,
               align=center, inner sep=2pt}
]
  \node[block] (p1) {\textbf{Phase 1}\\ ESE--DEC loop\\ + sticky fusion};
  \node[block, right=of p1] (p2) {\textbf{Phase 2}\\ ByT5 on $k^\star$\\ form $\mathcal{A}_{k^\star}$};
  \node[term, below=of p1] (succ) {Success};
  \node[term, below=of p2] (fail) {Fail};

  \draw[->,thick] ([yshift=2mm]p1.east) --
        node[above,font=\scriptsize]{stall}
        ([yshift=2mm]p2.west);
  \draw[->,thick] ([yshift=-2mm]p2.west) --
        node[below,font=\scriptsize]{resume}
        ([yshift=-2mm]p1.east);
  \draw[->,thick] (p1) --
        node[right,font=\scriptsize]{ \black{all accepted}}
        (succ);
  \draw[->,thick] (p2) --
        node[right,font=\scriptsize]{budget exhausted}
        (fail);
\end{tikzpicture}
\vspace{-0.5em}
\caption{Phase schedule of Sem-IR.}
\label{fig:phase-schedule}
\vspace{-1.75em}
\end{figure}

{The receiver alternates between two phases. \emph{Phase 1}
runs the ESE--DEC loop until
\black{the BP hard decisions stall}. \emph{Phase 2} invokes ByT5
on the lowest-index  \black{unaccepted} segment $k^\star$ to produce a
bit-level activation set $\mathcal{A}_{k^\star}$ and frozen
semantic LLRs, and
returns to Phase 1. From the moment a non-empty $\mathcal{A}_k$ is
created, every subsequent BP pass within Phase 1 reapplies the
fusion \eqref{eq:fusion} on that segment before the extrinsic is
fed back to the ESE; we refer to this as \emph{sticky fusion} and
detail it in Section~\ref{sec:sticky_fusion}. The two phases
alternate until either all segments are \black{accepted} or a per-segment
budget is exhausted, as illustrated in
Fig.~\ref{fig:phase-schedule}.
}

\subsubsection{NOMA Convergence Phase}

The receiver runs the ESE--DEC loop of
Section~\ref{sec:system_model} until
\black{the BP hard decisions have remained unchanged}
for $W_{\mathrm{stall}}$ consecutive outer
iterations or $t$ reaches $T_{\max}$, whichever occurs first.
{Whenever the activation set $\mathcal{A}_k$ of any segment is
non-empty, the sticky fusion \eqref{eq:fusion} is reapplied to
$\boldsymbol{L}_{\mathrm{APP},k}^{(t)}$ at every BP pass before the
decoder extrinsic \eqref{eq:dec_ext} is sent back to the ESE; the
detailed mechanism is described in
Section~\ref{sec:sticky_fusion}.}
The former {exit condition signals} a
decoding plateau; the latter {signals}
exhaustion of the iteration budget.
\black{At a stopping or prefix-selection decision, parity-valid
segments are evaluated by \eqref{eq:semantic_acceptance} to update
$\Omega_{\mathrm{acc}}^{(t)}$.}
\black{Decoding terminates successfully only if
$\Omega_{\mathrm{acc}}^{(t)}=\{1,\ldots,K\}$.}
\black{
\subsubsection{ByT5 Injection Phase}
Otherwise, let $k^\star\triangleq\min\{k:k\notin
\black{\Omega_{\mathrm{acc}}^{(t)}}\}$ be the lowest-index
 \black{unaccepted}
segment. We invoke ByT5 once on $k^\star$ to compute
$\mathcal{A}_{k^\star}$ and the frozen
$\{\widetilde{L}^{\mathrm{sem}}_{k^\star,\ell}\}_{\ell\in
\mathcal{A}_{k^\star}}$ via
\eqref{eq:bit_sem_llr}--\eqref{eq:active_set}. The receiver then
returns to the NOMA Convergence Phase{; from this
point on, the activation $\mathcal{A}_{k^\star}$ participates in
the sticky fusion alongside the activations of all previously
injected segments. If the resumed NOMA phase brings $k^\star$ into
$\black{\Omega_{\mathrm{acc}}^{(t)}}$, the next time this phase is
triggered it will target the new lowest-index
 \black{unaccepted} segment;
otherwise, $k^\star$ remains the target and a further ByT5
invocation may be issued, subject to the budget below}.
}
{
\subsubsection{Per-Segment Budget and Early Termination}
Each segment is allowed at most $N_{\mathrm{sem}}$ ByT5 invocations.
{Each invocation overwrites $\mathcal{A}_{k^\star}$
and $\{\widetilde{L}^{\mathrm{sem}}_{k^\star,\ell}\}$ with values
freshly computed from the current
$\hat{\boldsymbol{\mu}}_{k^\star}^{(t)}$, so successive attempts on
$k^\star$ are independent.}
If $k^\star$ has not joined
\black{$\Omega_{\mathrm{acc}}^{(t)}$} after all
$N_{\mathrm{sem}}$ attempts, the entire sentence is declared
unrecoverable and decoding terminates. This rule prevents repeated
low-quality injections on a stuck segment and bounds the worst-case
number of ByT5 forward passes per sentence by $N_{\mathrm{sem}} K$.
}
\subsubsection{Sticky Fusion Across Iterations} \label{sec:sticky_fusion}
{A ByT5 invocation on $k^\star$ is said to be \emph{successful}
if, in the immediately following NOMA Convergence Phase, segment
$k^\star$ joins
\black{$\Omega_{\mathrm{acc}}^{(t)}$}.} Once such a
successful invocation occurs, $\mathcal{A}_{k^\star}$ and the
frozen values $\{\widetilde{L}^{\mathrm{sem}}_{k^\star,\ell}\}
_{\ell\in\mathcal{A}_{k^\star}}$ are retained, and the fusion
\eqref{eq:fusion} is reapplied at every subsequent outer iteration
until the receiver terminates. After each BP pass produces a fresh
$\boldsymbol{L}_{\mathrm{APP},k}^{(t)}$, for every segment $k$ with
$\mathcal{A}_{k}\neq\emptyset$ the entries indexed by $\mathcal{A}_{k}$
are overwritten by \eqref{eq:fusion} before the decoder extrinsic
\eqref{eq:dec_ext} is computed and re-interleaved back to the ESE.

{
This persistence is necessary because the ESE forms its
soft-symbol estimates \eqref{eq:mean}--\eqref{eq:var} jointly across
all $K$ layers. A later ByT5 injection on $k^\star+1$ alters the
extrinsic feedback of layer $k^\star+1$ and, through the ESE, also
perturbs the input of layer $k^\star$. Without the sticky
re-injection, an already-\black{accepted} $k^\star$
may drift back out of
\black{$\Omega_{\mathrm{acc}}^{(t)}$}.
}

\subsubsection{Overall Algorithm}
The full procedure is summarized in
Algorithm~\ref{alg:sem_idma}.
\vspace{-1em}
\begin{algorithm}[h]
\small
\caption{semantic-aided iterative receiver (Sem-IR)}
\label{alg:sem_idma}
\begin{algorithmic}[1]
\REQUIRE $\{r(j)\}_{j=1}^{n}$; $T_{\max}$, $T_{\mathrm{bp}}$,$ W_{\mathrm{stall}},
  N_{\mathrm{sem}}, \alpha$, $\gamma, \tau_{\mathrm{avg}}$,$\tau_{\min}$
\ENSURE Decoded $\hat{\boldsymbol{\mu}}$ or \textsc{Fail}
\STATE Initialize $t\!\leftarrow\!0$,
       \black{$\Omega_{\mathrm{acc}}\!\leftarrow\!\emptyset$},
       $\mathcal{A}_k\!\leftarrow\!\emptyset$,
       $n_{\mathrm{sem}}(k)\!\leftarrow\!0$ for all $k$
\REPEAT
  \STATE \textit{// Phase 1: NOMA convergence (with sticky semantic injection)}
  \REPEAT
    \STATE $t\leftarrow t+1$
    \STATE Run one ESE--DEC pass via \eqref{eq:ese}--\eqref{eq:var}
    \STATE \textbf{For every} $k$ \textbf{with}
           $\mathcal{A}_k\neq\emptyset$: overwrite
           $\boldsymbol{L}_{\mathrm{APP},k}^{(t)}$ on $\mathcal{A}_k$
           via the sticky fusion \eqref{eq:fusion} \emph{before}
           computing \eqref{eq:dec_ext} and re-interleaving to the ESE
    \STATE \black{Update the BP hard decisions and their syndromes}
  \UNTIL{
         \black{the BP hard decisions remain unchanged}
         for $W_{\mathrm{stall}}$
         iterations \textbf{or} $t=T_{\max}$}
  \STATE \black{At a stopping or prefix-selection decision, evaluate
         \eqref{eq:semantic_acceptance} and update $\Omega_{\mathrm{acc}}$}
  \IF{
      \black{$\Omega_{\mathrm{acc}}=\{1,\ldots,K\}$}}
    \STATE \textbf{break}
  \ENDIF
  \STATE \textit{// Phase 2: ByT5 invocation on $k^\star$}
  \STATE $k^\star\leftarrow\min\{k\notin
         \black{\Omega_{\mathrm{acc}}}\}$
  \IF{$n_{\mathrm{sem}}(k^\star)=N_{\mathrm{sem}}$}
    \RETURN \textsc{Fail}
  \ENDIF
  \STATE Compute byte posteriors \eqref{eq:byte_post} on $k^\star$;
         marginalize to $\{L^{\mathrm{sem}}_{k^\star,\ell}\}$
         via \eqref{eq:bit_sem_logpost}--\eqref{eq:bit_sem_llr}
  \STATE Construct $\mathcal{A}_{k^\star}$ via \eqref{eq:active_set}
         and freeze $\widetilde{L}^{\mathrm{sem}}_{k^\star,\ell}
         \leftarrow L^{\mathrm{sem}}_{k^\star,\ell}$ for
         $\ell\in\mathcal{A}_{k^\star}$
          \STATE $n_{\mathrm{sem}}(k^\star)\leftarrow
         n_{\mathrm{sem}}(k^\star)+1$
\UNTIL{$t=T_{\max}$}
\RETURN $\hat{\boldsymbol{\mu}}$ if
       \black{$\Omega_{\mathrm{acc}}=\{1,\ldots,K\}$},
       else \textsc{Fail}
\end{algorithmic}
\end{algorithm}

\vspace{-0.8em}
\section{Simulation Results} \label{sec:simulation}
\vspace{-0.5em}

\vspace{-0.3em}
\subsection{System Setup}

\vspace{-0.3em}

\subsubsection{Dataset and Training}
We evaluate Sem-IR on natural-language sentences drawn
from the English Wikipedia corpus, retaining $5\!\times\!10^{4}$
sentences whose lengths fall in $[121,128]$ characters and splitting
them sentence-disjointly into training, validation, and test sets at
a ratio of $80\!:\!10\!:\!10$.

The byte-level prior of Section~\ref{sec:byt5} is initialised from
the public ByT5-Small checkpoint and augmented with the
classification head defined in
\eqref{eq:byt5_input}--\eqref{eq:byte_post}. Training samples are
collected by running the full NOMA simulation under the present
setup over the $\mathrm{SNR}$ range $4.0$--$8.0$\,dB in $0.5$\,dB
steps; from each pass we extract a training instance
$(\boldsymbol{\mu}_{1:k-1},\hat{\boldsymbol{\mu}}_k)$ in which $k$
is the lowest-index segment failing the LDPC parity check and
$\hat{\boldsymbol{\mu}}_k$ is the byte form of its BP
information-bit hard decisions, yielding on the order of
$1\!\times\!10^{5}$ instances in total. The model is
fine-tuned by minimising \eqref{eq:ce_loss} with AdamW
(learning rate $2\!\times\!10^{-4}$, weight decay $0.01$) under a
cosine schedule with $10\%$ warm-up.

\subsubsection{Channel Code, Power Allocation, and Receiver Parameters}
The sentence is partitioned into $K=8$ users, each with
$L=16$ bytes ($M=128$ information bits) encoded by a rate-$R=0.1$
QC-LDPC code into a codeword of length $n=1280$. The $K$ encoded
layers are interleaved by independent random permutations
$\{\pi_k\}$, BPSK modulated, and superimposed under the geometric
power profile \eqref{eq:power} with ratio $\rho=1.26$. The total
transmit power is normalised to $\sum_{k=1}^{K}p_k=1$. We report
performance against the channel signal-to-noise ratio
$\mathrm{SNR}=1/\sigma^{2}$ defined in
Section~\ref{sec:system_model}.

The iterative receiver of Section~\ref{sec:receiver} is configured
as follows: maximum outer iterations $T_{\max}=80$,
$T_{\mathrm{bp}}=80$ BP rounds, stall window
$W_{\mathrm{stall}}=4$, per-segment ByT5 budget
$N_{\mathrm{sem}}=3$, fusion mixing weight
$\alpha=0.9$, bit-level confidence threshold
$\gamma=8$, and damping factor
$\beta=0.5$.
\black{For language-plausibility acceptance, the average and worst-byte
thresholds are $\tau_{\mathrm{avg}}=-0.1$ and
$\tau_{\min}=-0.7$, respectively.}

\subsubsection{Benchmarks}
The proposed scheme, denoted Sem-IR, is compared against:
\begin{itemize}
\setlength{\itemsep}{0pt}
\setlength{\parskip}{0pt}
\setlength{\parsep}{0pt}
\item \textbf{TDMA}: $K$ orthogonal slots over the same total duration;
each segment uses a rate-$KR=0.8$ LDPC code and one BP decode.
\item \textbf{TDMA + 1 ByT5}: TDMA followed by one ByT5 pass on failed
segments using the decoded prefix.
\item \textbf{NOMA}: the same ESE--DEC receiver and parameters as
Sem-IR, with $\mathcal{A}_k=\emptyset$.
\item \textbf{NOMA + 1 ByT5}: NOMA followed by one ByT5 pass on failed
segments using the decoded prefix.
\end{itemize}
All schemes are matched in total channel uses, total information
bits per sentence, and average transmit energy. We report block error rate (BLER), defined as the fraction of
sentences containing at least one residual bit error after decoding.

\vspace{-0.5em}
\subsection{Complexity Analysis}
\label{sec:complexity}

\vspace{-0.3em}

{Let $T_b$ be the BP rounds per LDPC decode and
$C_{\mathrm{B}}=\mathcal{O}(z^{2}d+zd^{2})$ the cost of one ByT5
forward pass, where $z=\mathcal{O}(KL)$ is the input length and $d$
is the encoder hidden dimension. Table~\ref{tab:complexity} reports
the per-sentence worst-case cost of the four schemes.}

\begin{table}[t]
\centering
\caption{Per-sentence worst-case complexity, where $T_b$ is the BP rounds per LDPC decode.}
\label{tab:complexity}
\renewcommand{\arraystretch}{1.15}
\footnotesize
\begin{tabular}{@{}lll@{}}
\toprule
\textbf{Scheme} & \textbf{ESE+BP} & \textbf{ByT5} \\
\midrule
TDMA           & $\mathcal{O}(K n_{\mathrm{ser}} T_b)$ & --- \\
TDMA + 1 ByT5  & $\mathcal{O}(K n_{\mathrm{ser}} T_b)$
               & $\mathcal{O}(K\,C_{\mathrm{B}})$ \\
NOMA           & $\mathcal{O}(T_{\max} K n T_b)$ & --- \\
NOMA + 1 ByT5  & $\mathcal{O}(T_{\max} K n T_b)$
               & $\mathcal{O}(K\,C_{\mathrm{B}})$ \\
Sem-IR         & $\mathcal{O}(T_{\max} K n T_b)$
               & $\mathcal{O}(N_{\mathrm{sem}} K\,C_{\mathrm{B}})$ \\
\bottomrule
\end{tabular}
\vspace{-1em}
\end{table}

The proposed {Sem-IR} is the most expensive of the five
schemes. Compared with {TDMA}, it pays a factor of $T_{\max}$
in the LDPC term for the iterative ESE--DEC loop and an additional
ByT5 term of up to $N_{\mathrm{sem}}K$ forward passes. Compared
with {TDMA + 1 ByT5} and {NOMA + 1 ByT5}, the
per-segment ByT5 budget grows from one invocation to
$N_{\mathrm{sem}}$. The sticky fusion \eqref{eq:fusion} contributes
only $\mathcal{O}(M)$ per BP pass and is negligible against the BP
and ByT5 terms.

The gap between {NOMA} and {TDMA} reflects the standard
complexity premium of NOMA-class receivers over orthogonal access:
joint detection across superimposed layers requires the iterative
ESE--DEC loop, whereas {TDMA} reduces to $K$ independent BP
decodes. This premium is intrinsic to non-orthogonal transmission
and is paid in exchange for the joint spectral-efficiency and
feedback gains exploited by the proposed scheme. The further
$\mathcal{O}(N_{\mathrm{sem}}K\,C_{\mathrm{B}})$ term in
{Sem-IR} amounts to at most $N_{\mathrm{sem}}K$ ByT5 forward
passes, whose wall-clock cost is set primarily by the available
GPU.

\vspace{-0.3em}
\subsection{Simulation Results}
\vspace{-0.3em}

\subsubsection{BLER}
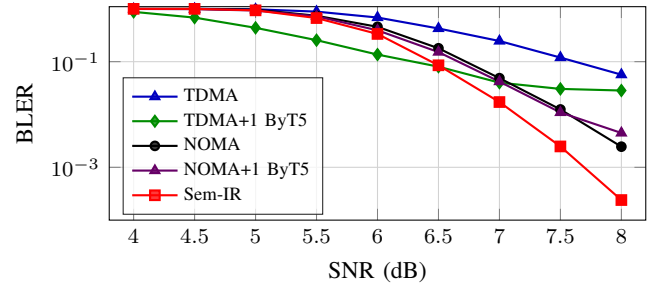
\begin{figure}[t]
\centering
\begin{tikzpicture}
\begin{axis}[
    width=0.98\linewidth,
    height=4.4cm,
    xlabel={SNR (dB)},
    ylabel={BLER},
    ymode=log,
    log basis y=10,
    xmin=3.8,
    xmax=8.2,
    ymin=1e-4,
    ymax=1.1,
    xtick={4,4.5,5,5.5,6,6.5,7,7.5,8},
    tick label style={font=\footnotesize},
    label style={font=\small},
    xmajorgrids,
    ymajorgrids,
    yminorgrids,
    major grid style={solid,gray!35},
    minor grid style={dotted,gray!25},
    legend style={
        at={(0.025,0.025)},
        anchor=south west,
        legend cell align=left,
        draw=white!15!black,
        font=\scriptsize,
        row sep=-1pt,
        fill opacity=0.78,
        text opacity=1
    }
]

\addplot[mark=triangle*, mark size=1.9pt, color=blue!75!black, thick] coordinates {
    (4.50, 1.00000000e+00)
    (5.00, 9.80392157e-01)
    (5.50, 8.86699507e-01)
    (6.00, 6.85518423e-01)
    (6.50, 4.25079702e-01)
    (7.00, 2.46132208e-01)
    (7.50, 1.19593382e-01)
    (8.00, 5.67536890e-02)

};
\addlegendentry{TDMA}

\addplot[mark=diamond*, mark size=2.0pt, color=green!55!black, thick] coordinates {
    (4.00, 8.77192982e-01)
    (4.50, 6.82926829e-01)
    (5.00, 4.37445319e-01)
    (5.50, 2.55427842e-01)
    (6.00, 1.36301681e-01)
    (6.50, 8.04289544e-02)
    (7.00, 4.00641026e-02)
    (7.50, 3.05157156e-02)
    (8.00, 2.84414107e-02)
};
\addlegendentry{TDMA+1 ByT5}

\addplot[mark=*, mark size=1.6pt, color=black, thick] coordinates {
    (4.00, 1.00000000e+00)
    (4.50, 1.00000000e+00)
    (5.00, 9.66183575e-01)
    (5.50, 7.40242261e-01)
    (6.00, 4.57038391e-01)
    (6.50, 1.79452669e-01)
    (7.00, 4.87646294e-02)
    (7.50, 1.24595066e-02)
    (8.00, 2.45817014e-03)
    
};
\addlegendentry{NOMA}

\addplot[mark=triangle*, mark size=1.9pt, color=violet!80!black, thick] coordinates {
    (4.00, 9.96028791e-01)
    (4.50, 9.86100769e-01)
    (5.00, 9.31496649e-01)
    (5.50, 7.22511790e-01)
    (6.00, 3.92405063e-01)
    (6.50, 1.51898734e-01)
    (7.00, 4.19458923e-02)
    (7.50, 1.09208240e-02)
    (8.00, 4.46760983e-03)
};
\addlegendentry{NOMA+1 ByT5}

\addplot[mark=square*, mark size=1.8pt, color=red, thick] coordinates {

(4,1)
(4.5,1)
    (5.00, 9.37500000e-01)
    (5.50, 6.69642857e-01)
    (6.00, 3.35008375e-01)
    (6.50, 8.53970965e-02)
    (7.00, 1.72087420e-02)
    (7.50, 2.48434860e-03)
    (8.00, 2.40000000e-04)
};
\addlegendentry{Sem-IR}

\end{axis}
\end{tikzpicture}
\vspace{-1em}
\caption{BLER versus SNR for Sem-IR and benchmark schemes}
\label{fig:selected-bler-comparison}
\vspace{-1em}
\end{figure}

{Fig.~\ref{fig:selected-bler-comparison} demonstrates BLER versus SNR
for the five schemes. At low SNR ($\leq\!5$\,dB) all schemes are
noise-limited. NOMA enters its waterfall near $6$\,dB and reaches
$2.5\!\times\!10^{-3}$ at $8$\,dB, already an order of magnitude
better than TDMA. Sem-IR further improvess the BLER performance, achieving $2.4\!\times\!10^{-4}$ at $8$\,dB,
a $10\times$ reduction over NOMA. NOMA + 1 ByT5 stays essentially
aligned with NOMA across the entire range, confirming that the
gain of Sem-IR comes from the
iterative loop between the language-model and channel-code
decoders rather than from a single language-model query.} TDMA + 1 ByT5 plateaus at high SNR, where its performance is bounded by the intrinsic prediction error rate of the language
model itself.

\begin{figure}[t]
\centering
\begin{tikzpicture}
\begin{axis}[
    width=0.98\linewidth,
    height=4.4cm,
    xlabel={SNR (dB)},
    ylabel={Avg. Correct Segments},
    xmin=5,
    xmax=8.2,
    ymin=4,
    ymax=8.25,
    xtick={4,4.5,5,5.5,6,6.5,7,7.5,8},
    ytick={0,1,2,3,4,5,6,7,8},
    tick label style={font=\footnotesize},
    label style={font=\small},
    xmajorgrids,
    ymajorgrids,
    major grid style={solid,gray!35},
    legend style={
        at={(0.975,0.025)},
anchor=south east,
        legend cell align=left,
        draw=white!15!black,
        font=\scriptsize,
        row sep=-1pt,
        fill opacity=0.78,
        text opacity=1
    }
]
\addplot[mark=square*, mark size=1.8pt, color=red, thick] coordinates {
    (5.00, 5.590625)
    (5.50, 6.631696)
    (6.00, 7.376884)
    (6.50, 7.885568)
    (7.00, 7.978833)
    (7.50, 7.996721)
    (8.00, 7.999410)
};
\addlegendentry{Sem-IR}

\addplot[mark=*, mark size=1.6pt, color=black, thick] coordinates {
    (4.00, 3.109456)
    (4.50, 4.034003)
    (5.00, 5.097543)
    (5.50, 6.257384)
    (6.00, 7.216679)
    (6.50, 7.736907)
    (7.00, 7.934227)
    (7.50, 7.984612)
    (8.00, 7.993671)
};
\addlegendentry{NOMA}

\addplot[mark=diamond*, mark size=1.9pt, color=violet!80!black, thick]  coordinates {
    (4.00, 3.126086)
    (4.50, 4.069000)
    (5.00, 5.149665)
    (5.50, 6.320427)
    (6.00, 7.255647)
    (6.50, 7.754778)
    (7.00, 7.936957)
    (7.50, 7.982130)
    (8.00, 7.990196)
};
\addlegendentry{NOMA+1 ByT5}

\addplot[mark=triangle*, mark size=1.9pt, color=blue!75!black, thick] coordinates {
    (4.00, 1.603070)
    (4.50, 3.173913)
    (5.00, 4.742087)
    (5.50, 6.015873)
    (6.00, 6.989130)
    (6.50, 7.311688)
    (7.00, 7.705455)
    (7.50, 7.856863)
    (8.00, 7.939690)
};
\addlegendentry{TDMA}

\end{axis}
\end{tikzpicture}
\vspace{-1em}
\caption{Average number of correctly recovered users versus SNR.}
\vspace{-1em}
\label{fig:avg-correct-segments}
\end{figure}
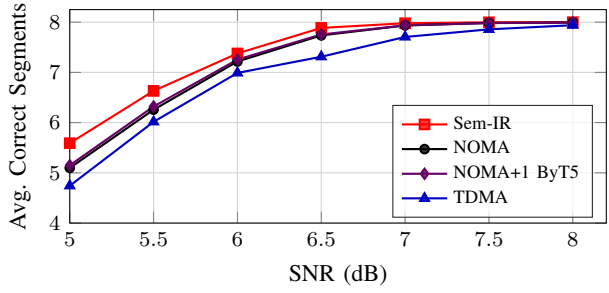

\subsubsection{Number of Correct Users}
Fig.~\ref{fig:avg-correct-segments} reports the average number
of correctly recovered users out of $K=8$. Sem-IR consistently
recovers more users than NOMA in the mid-SNR range: at $5$\,dB
it decodes $5.59$ users against $5.10$ for NOMA and $4.74$ for
TDMA, and the advantage persists up to $6.5$\,dB. NOMA + 1 ByT5
overlaps NOMA almost exactly ($5.15$ versus $5.10$ at $5$\,dB). The results confirm that a single open-loop language-model query offers
no meaningful gain over the iterative ESE--DEC receiver. In the multi-user setting, the unrecovered users sit under
heavy residual interference, and the short prefix from the few
recovered users limits the language model's prediction accuracy.

\subsubsection{Effect of Confidence Threshold}
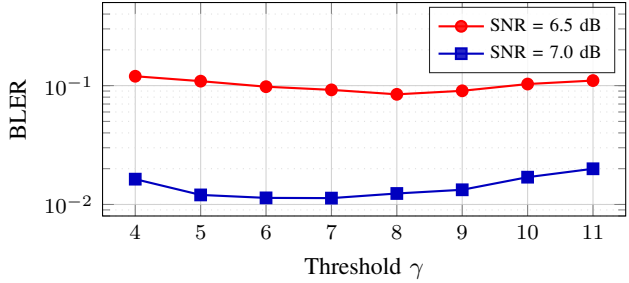
\begin{figure}[t]
\centering
\begin{tikzpicture}
\begin{axis}[
    width=0.96\linewidth,
    height=4.4cm,
    xlabel={Threshold $\gamma$},
    ylabel={BLER},
    ymode=log,
    log basis y=10,
    xmin=3.5,
    xmax=11.5,
    ymin=0.8e-2,
    ymax=5e-1,
    xtick={4,5,6,7,8,9,10,11},
    tick label style={font=\footnotesize},
    label style={font=\small},
    xmajorgrids,
    ymajorgrids,
    yminorgrids,
    major grid style={solid,gray!35},
    minor grid style={dotted,gray!25},
    legend style={
        at={(0.98,0.98)},
        anchor=north east,
        legend cell align=left,
        draw=white!15!black,
        font=\scriptsize,
        fill opacity=0.85,
        text opacity=1
    }
]

\addplot[
    mark=*,
    mark size=2.0pt,
    color=red,
    thick
] coordinates {
    (4.00, 1.19904077e-01)
    (5.00, 1.09018062e-01)
    (6.00, 9.79259260e-02)
    (7.00, 9.20810313e-02)
    (8.00, 8.44594595e-02)
    (9.00, 9.04159132e-02)
    (10.00, 1.03092784e-01)
    (11.00, 1.10263158e-01)
};
\addlegendentry{SNR = 6.5 dB}

\addplot[
    mark=square*,
    mark size=2.0pt,
    color=blue!75!black,
    thick
] coordinates {
    (4.00, 1.63300000e-02)
    (5.00, 1.20300000e-02)
    (6.00, 1.13600000e-02)
    (7.00, 1.13100000e-02)
    (8.00, 1.23700000e-02)
    (9.00, 1.32900000e-02)
    (10.00, 1.69700000e-02)
    (11.00,1.99700000e-02)
};
\addlegendentry{SNR = 7.0 dB}

\end{axis}
\end{tikzpicture}
\vspace{-1em}
\caption{BLER versus semantic confidence threshold $\gamma$ under different SNRs.}
\vspace{-1.5em}
\label{fig:gamma-sweep-bler}
\end{figure}

Fig.~\ref{fig:gamma-sweep-bler} shows BLER as a function of the
semantic confidence threshold $\gamma$ at two SNRs. Both curves
exhibit a clear U-shape: a small $\gamma$ admits low-confidence
ByT5 predictions that mislead the BP decoder, while a large
$\gamma$ discards too many semantic LLRs and erodes the gain
from the language model. The optimum sits at $\gamma\!=\!8$ for
SNR$\,=\,6.5$\,dB and at $\gamma\!=\!7$ for SNR$\,=\,7.0$\,dB,
with a roughly $1.4\!\times$ BLER spread across the swept range.
The optimum shifts down with SNR because cleaner channel evidence
reduces the need for aggressive filtering of the semantic prior.

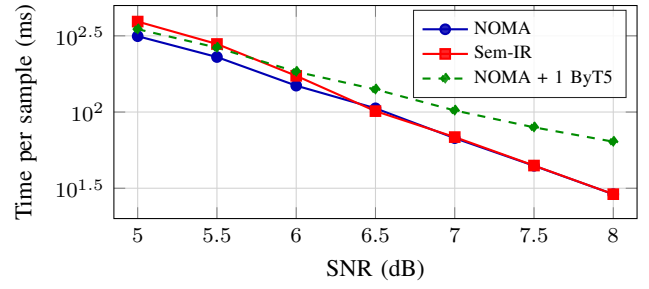
\begin{figure}[t]
\centering
\begin{tikzpicture}
\begin{axis}[
    width=0.96\linewidth,
    height=4.4cm,
    xlabel={SNR (dB)},
    ylabel={Time per sample (ms)},
    ymode=log,
    log basis y=10,
    xmin=4.85,
    xmax=8.15,
    ymin=2.0e1,
    ymax=5.0e2,
    xtick={5,5.5,6,6.5,7,7.5,8},
    tick label style={font=\footnotesize},
    label style={font=\small},
    xmajorgrids,
    ymajorgrids,
    yminorgrids,
    major grid style={solid,gray!35},
    minor grid style={dotted,gray!25},
    legend style={
        at={(0.98,0.98)},
        anchor=north east,
        legend cell align=left,
        draw=white!15!black,
        font=\scriptsize,
        row sep=-1pt,
        fill opacity=0.85,
        text opacity=1
    }
]

\addplot[
    mark=*,
    mark size=1.8pt,
    color=blue!70!black,
    thick
] coordinates {
    (5.0, 314.6)
    (5.5, 229.6)
    (6.0, 149.0)
    (6.5, 105.8)
    (7.0, 67.4)
    (7.5, 44.4)
    (8.0, 28.9)
};
\addlegendentry{NOMA}

\addplot[
    mark=square*,
    mark size=1.8pt,
    color=red,
    thick
] coordinates {
    (5.0, 392.4)
    (5.5, 279.6)
    (6.0, 172.9)
    (6.5, 101.4)
    (7.0, 68.5)
    (7.5, 44.56)
    (8.0, 28.92)
};
\addlegendentry{Sem-IR}

\addplot[
    mark=diamond*,
    mark size=2.0pt,
    color=green!55!black,
    thick,
    dashed
] coordinates {
    (5.0, 349.6)
    (5.5, 264.6)
    (6.0, 184.0)
    (6.5, 140.8)
    (7.0, 102.4)
    (7.5, 79.4)
    (8.0, 63.9)
};
\addlegendentry{NOMA + 1 ByT5}

\end{axis}
\end{tikzpicture}
\vspace{-0.5em}
\caption{Average decoding time per sample under different receiver strategies.}
\vspace{-1.25em}
\label{fig:complexity-time}
\end{figure}

\subsubsection{Decoding Latency}
Fig.~\ref{fig:complexity-time} reports the average decoding time
per sentence, measured on an NVIDIA RTX 5090 GPU. All schemes
accelerate with SNR as fewer outer iterations are needed. The
overhead of Sem-IR over NOMA shrinks rapidly: from $25\%$ at
$5$\,dB ($392$\,ms vs $315$\,ms) to under $2\%$ at $7$\,dB and to
negligible at $8$\,dB ($28.9$\,ms for both). At high SNR, the
ESE--DEC loop converges in the first few outer iterations, so
Sem-IR's ByT5 budget is rarely triggered. NOMA + 1 ByT5, in
contrast, invokes the language model once per sentence regardless
of channel quality, paying a roughly constant $35$\,ms overhead
that becomes the dominant cost at $8$\,dB ($64$\,ms vs $29$\,ms
for NOMA). Sem-IR thus pays for the language model only when the
iterative receiver actually needs it.

{
\vspace{-0.5em}
\section{Conclusion} \label{sec:conclusion}
\vspace{-0.3em}
This paper proposed Sem-IR, a semantic-aided iterative
receiver that closes the loop between a byte-level language model
and the physical-layer iteration in LDPC-encoded uplink
non-orthogonal transmission of natural language. Byte posteriors
from a fine-tuned ByT5 are marginalized to bit-level LLRs, gated
by confidence, and convex-combined with the LDPC a-posteriori
LLRs; the injected LLRs are held across subsequent iterations to
anchor repaired users under a per-user invocation budget.
Simulations show that Sem-IR outperforms TDMA and the same NOMA
receiver without semantic feedback in block error rate.
}

\bibliographystyle{IEEEtran}
\bibliography{references}
\end{document}